\documentclass{PoS}

\usepackage{amsmath}

\title{Dielectronic Recombination Rates in Astrophysical Plasmas}

\ShortTitle{Dielectronic Recombination Rates In Astrophysical Plasmas}

\abstract{In this work we introduce a new expression of the plasma Dielecronic Recombination (DR) rate as a function of the temperature, derived assuming a small deformation of the Maxwell-Boltzmann distribution and containing corrective factors, in addition to the usual exponential behaviour, caused by non-linear effects in slightly non ideal plasmas. We then compare the calculated DR rates with the experimental DR fits in the low temperature region.}

\author{\speaker{Fatima~Bachari}\\
Politecnico di Torino - Dipartimento di Fisica, Italy\\
E-mail: \email{fatima.bachari@polito.it}}

\author{Fabrizio~Ferro\\
Politecnico di Torino - Dipartimento di Fisica and INFN - Sezione di Torino, Italy\\
E-mail: \email{fabrizio.ferro@polito.it}}

\author{Giancarlo~Maero\\
Gesellschaft f\"{u}r Schwerionenforschung (GSI) - Darmstadt, Germany\\
E-mail: \email{g.maero@gsi.de}}

\author{Piero~Quarati\\
Politecnico Torino - Dipartimento di Fisica and INFN - Sezione di Cagliari, Italy\\
E-mail: \email{piero.quarati@polito.it}}

\FullConference{International Symposium on Nuclear Astrophysics - Nuclei in the Cosmos - IX\\
25-30 June 2006\\
CERN}

\begin{document}
\section{Introduction}
Dielectronic Recombination (DR) plays a central role in astrophysics. It is the most important electron-ion recombination process in photoionized cosmic plasmas. Moreover, DR rates are used to model different astrophysical systems as stellar and solar coronae and the intergalactic medium~\cite{Mul1999,Sav2000}. Theoretical and experimental uncertainties in the low temperature DR rates can dramatically change the values of parameters describing the properties of the cosmic plasmas and the interpretation of data from satellites on cosmic systems.

Actually, the quality and accuracy of the laboratory experiments on DR processes with colliding beams of electrons and ions are very high and
atomic codes to evaluate transition amplitudes rather sophisticated~\cite{Sch2001,Sch2006}. The transfer of knowledge obtained from laboratory measurements of the DR rates to astrophysical plasma modelling is based on the assumption that the electron beam of the cooling device can be described with a Maxwellian distribution with energy spread as low as about 10 meV (experimentally verified) and that the astrophysical plasma be in a state of global thermodynamical equilibrium or in a LTE with a Maxwellian distribution over a wide range of temperature from $10^{-3}\,\mathrm{eV}$ to $10^{6}\,\mathrm{eV}$.

On the contrary, signals of deviations from LTE in astrophysical plasmas have been observed and reported in the past~\cite{Sho1969,Col1989} and, as a very recent result, astrophysical X-ray sources active in stellar coronae, active galactic nuclei, X-ray binaries, supernovae and after-glows from gamma-ray bursts indicate the presence of a departure from Maxwellian distribution of the particles in many different types of plasmas~\cite{Guo2004}.

We have recently shown that, in electron cooling devices, generalized electron distributions that differ from the Maxwellian one in the low energy part, due to subdiffusion of electrons and ions, with depleted energy tails, may account for the observed enhancement in Radiative Recombination (RR) and that the existence of a cut-off in the momentum distribution (as the one introduced in astrophysics by Spitzer in the past~\cite{Spi1940}) could have important consequences on the X-ray spectra emitted during RR~\cite{Mae2006}.

In this work we introduce a new expression of the plasma DR rate as a function of the temperature, derived assuming a small deformation of the Maxwell distribution and containing factors that modify the usual, well known exponential behaviour. Deviations are caused by presence of correlations among particles in non ideal plasmas, of random electric microfields distribution and of random forces whose effects are non linear~\cite{Fer2005,Fer2006}. All these effects are taken into account by means of an appropriate parameter $q$, which is the entropic parameter of the nonextensive Tsallis thermostatistics~\cite{Boo2005,Gel2004}. Among many generalised statistics, this is the one (together with Renyi statistics) which introduces a cut-off in the energy distribution when particles sub-diffuse ($q<1$) .

We apply this new DR rate to the study of the $\mathrm{C}^{3+}$ case, as for carbon ions there exist experimental data and theoretical calculations we can use in our semi-phenomenological treatment and may compare with our results. We want to show what is the effect of the deformation on the plasma DR rate and calculate, in particular, the rate at low temperature. The same application will be extended elsewhere to the DR of Fe ions exploiting the experimental and theoretical results of Ref.~\cite{Sav2006}.

\section{The dielectronic recombination rates}
The general formula for a reaction rate with a Maxwell-Boltzmann statistics, by definition, reads
\begin{equation}
\alpha(k_{B} T)=\langle\sigma v\rangle=4\pi\left(\frac{\mu}{2\pi k_{B} T}\right)^{3/2}\int_{0}^{+\infty}v^{3}\sigma(v)\exp\left(-\frac{\mu v^{2}}{2k_{B} T}\right)\,\mathrm{d}v\, ,\label{Maxwellian reaction rate}
\end{equation}
where $\mu$ is the ion-electron reduced mass, $T$ is the temperature of the system, and $\sigma(v)$ is the interaction cross section.

For a given DR resonance $i$ located at $E_{\mathrm{res},i}$ and characterized by the width $\Gamma_{i}$ and the strength $\bar{\sigma}_{i}$, the corresponding cross section can be written as
\begin{displaymath}
\sigma_{i}(E)=\frac{\bar{\sigma}_{i}}{\pi}\frac{E_{\mathrm{res},i}}{E}\frac{\Gamma_{i}/2}{(E-E_{\mathrm{res},i})^{2}+(\Gamma_{i}/2)^{2}}\, .
\end{displaymath}

By using the Clayton-Tsallis distribution for small deformation, if the distribution has a cut-off at $E=k_{B} T/(2\delta)$ (for $\delta>0$), with $\delta=(1-q)/2$~\cite{Cor1999}, the reaction rate~\ref{Maxwellian reaction rate} transforms into
\begin{equation}
\alpha_{(\delta,\gamma)}^{\mathrm{DR}}(k_{B} T)=\underset{i}\sum\alpha_{i_{(\delta,\gamma)}}^{\mathrm{DR}}(k_{B} T)\, ,\label{our plasma rate}
\end{equation}
with
\begin{eqnarray}
\alpha_{i_{(\delta,\gamma)}}^{\mathrm{DR}}(k_{B} T)&=&\left(\frac{2}{\pi k_{B} T}\right)^{\frac{3}{2}}\frac{\bar{\sigma}_{i}\,E_{\mathrm{res},i}\,\Gamma_{i} c}{2\sqrt{\mu c^{2}}}\left(1+\frac{15}{4}\delta\right)\times\nonumber\\
& &\int_{0}^{\frac{k_{B} T}{2\delta}}\mathrm{d}E\,\frac{1}{(E-E_{\mathrm{res},i})^{2}+(\Gamma_{i}/2)^{2}}\cdot\exp\left[-\frac{E}{k_{B} T}-\delta \left(\frac{E}{k_{B} T}\right)^{2}\right]\, .\label{rate to be calculated}
\end{eqnarray}

Working out the integral in Eq.~\ref{rate to be calculated}, the general expression of the rate~\ref{our plasma rate} becomes
\begin{eqnarray}
\alpha_{(\delta,\gamma)}^{\mathrm{DR}}(k_{B} T)&=&
\sqrt{\frac{2}{\pi}}\left(1+\frac{15}{4}\delta\right)\underset{i}\sum\frac{\bar{\sigma}_{i}\,E_{\mathrm{res},i}\,c}{(k_{B} T)^{\frac{3}{2}}\sqrt{\mu c^{2}}}\left[1-\frac{\delta}{2}\left(\frac{\gamma_{i}}{k_{B} T}\right)^{2}\right]\times\nonumber\\
& &\exp\left[-\frac{E_{\mathrm{res},i}}{k_{B} T}\left(1+\delta\,\frac{E_{\mathrm{res},i}}{k_{B} T}-\frac{\gamma_{i}^{2}}{4E_{\mathrm{res},i}
k_{B} T}\right)\right]\times\nonumber\\
& &\exp\left[-\delta\left(\frac{\gamma_{i}}{k_{B} T}\right)^{2}\left(\frac{\gamma_{i}^{2}}{4\left(k_{B} T\right)^{2}}-
\frac{E_{\mathrm{res},i}}{k_{B} T}\right)\right]\times\nonumber\\
& &\left[\mathrm{Erf}\left(b\sqrt{\frac{1}{4a}}+\frac{k_{B} T\sqrt{4a}}{4\delta}\right)-
\mathrm{Erf}\left(b\sqrt{\frac{1}{4a}}\right)\right]\, ,\label{non-Maxwellian rate}
\end{eqnarray}
with
\begin{equation}
\begin{cases}
b\sqrt{\displaystyle{\frac{1}{4a}}}=\left(\displaystyle{\frac{\gamma_{i}}{2k_{B} T}}-\displaystyle{\frac{E_{\mathrm{res},i}}{\gamma_{i}}}\right)
\cdot\left[1-\displaystyle{\frac{\delta}{2}}\left(\displaystyle{\frac{\gamma_i}{k_{B} T}}\right)^{2}\right]\\
\displaystyle{\frac{k_{B} T\sqrt{4a}}{4\delta}}=\displaystyle{\frac{k_{B}T}{2\delta}\frac{1}{\gamma_{i}}}\left[1+\frac{\delta}{2}\left(\displaystyle{\frac{\gamma_{i}}{k_{B} T}}\right)^{2}\right]\nonumber
\end{cases}
\end{equation}
and $\gamma_{i}=\Gamma_{i}\sqrt{\pi}/2$, under the condition $\delta<2(k_{B} T/\gamma_{i})^{2}$.

With a pure Maxwellian distribution, for any $k_{B}T$, the energy
tail contains particles of high energy. If the lowest resonance
level contributing to the DR rate lies at $E_{\mathrm{res},i}>k_{B}
T$, the factor $\exp(-E_{\mathrm{res},i}/k_{B} T)$ that goes to zero
for low $k_{B} T$ must balance with the factor $(k_{B} T)^{-3/2}$.

From the general expression~\ref{non-Maxwellian rate}, we can see
that all resonances contribute to the rate when the distribution is
Maxwellian (recovered with $\delta=0$).

If the distribution has a cut-off, the particles with $E>k_{B}
T/(2\delta)$ are absent and the factor $(k_{B} T)^{-3/2}$ must be
balanced with $\exp\left[-\frac{E_{\mathrm{res},i}}{k_{B}
T}+\delta\left(\frac{E_{\mathrm{res},i}}{k_{B}
T}\right)^{2}\right]$.

The Maxwellian and the slightly deformed distribution intersect at
$E_{\mathrm{int}}=2k_{B} T$. Let us consider a resonance at
$E_{\mathrm{res},i}$ with a negligible width $\Gamma_{i}$. If
$E_{\mathrm{res},i}<E_{\mathrm{int}}$, the rate
$\alpha_{i}^{\mathrm{DR}}(k_{B} T)$ of Eq.~\ref{rate to be
calculated} corresponding to this resonance increases when
$\delta>0$ and decreases when $\delta<0$. If on the contrary
$E_{\mathrm{res},i}>E_{\mathrm{int}}$, the rate
$\alpha_{i}^{\mathrm{DR}}(k_{B} T)$ of this resonance increases with
respect to the Maxwellian rate if $\delta<0$ and decreases if
$\delta>0$.

If we do not neglect the width $\Gamma_{i}$ of the resonance, the
factor $\exp\left[-\frac{E_{\mathrm{res},i}}{k_{B}
T}\left(1-\frac{\gamma_{i}^{2}}{4E_{\mathrm{res},i}k_{B}
T}\right)\right]$ can have a significant role. If
$\gamma_{i}^{2}<4E_{\mathrm{res},i}\,k_{B} T$ the rate is depleted,
otherwise when $\gamma_{i}^{2}>4E_{\mathrm{res},i}\,k_{B} T$ the
rate is enhanced.

Using the non-Maxwellian $q$-distribution, only resonance levels
below the cut-off contribute to the rate. In this case the important
corrective factor is $\exp\left[-\frac{E_{\mathrm{res},i}}{k_{B}
T}\left(1+\delta\frac{E_{\mathrm{res},i}}{k_{B} T}
-\frac{\gamma_{i}^{2}}{4E_{\mathrm{res},i}}{k_{B} T}\right)\right]$,
and the rate is less depleted with respect to the previous case as
$k_{B} T$ increases.

An astrophysical plasma at $10^{5}-10^{6}\,\mathrm{K}$ has a value
of plasma parameter that strongly differs from the value at
$T=10^{2}-10^{3}\,\mathrm{K}$ if the electron density does not
diminish of the factor $10^{9}$. The correlations among the
particles are different in the two extreme cases and therefore the
plasma does not preserve its Maxwellian behaviour. We must consider
at each value of the plasma parameter a given deviation from the
ideal case and thus a different degree of non-extensivity described
by a suitable value of the parameter $\delta$. For instance, above
$10^{5}\,\mathrm{K}$, we assume $\delta=0$, because the system is
Maxwellian, having a plasma parameter much smaller than one, while
for $T<10^{3}\,\mathrm{K}$, we could assume $\delta=0.2$.

\begin{figure}
\begin{center}

\includegraphics[width=0.9\textwidth]{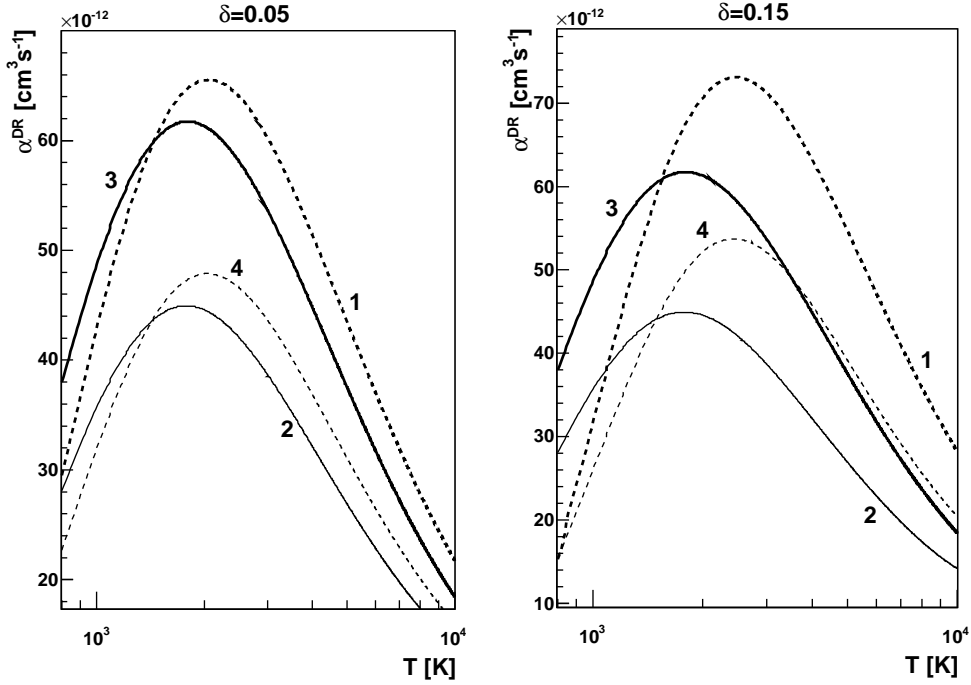}

\caption{Plots of dielectronic recombination rates against the plasma temperature ($\delta=0.05$ left, and $\delta=0.15$ right). The four curves refer to: 1) the non-Maxwellian recombination rate $\alpha^{\mathrm{DR}}_{(\delta,\gamma)}(k_{B} T)$ with 21 resonances (thick dashed line); 2) the fit from Schippers~{\it et al.}~\cite{Sch2001}, used for comparison with our results (thin solid line); 3) the Maxwellian recombination rate $\alpha^{\mathrm{DR}}_{(\delta,\gamma)}(k_{B} T)$ with 21 resonances and $\delta=0$ (thick solid line); 4) the non-Maxwellian fit $\alpha_{\mathrm{fit}}^{\mathrm{DR}}(k_{B} T)$ (thin dashed line).}\label{recombination plot}

\end{center}
\end{figure}

We chose to restrict ourselves to a temperature range between
$T=0.8\cdot 10^{3}\,\mathrm{K}$ and $T=10^{4}\,\mathrm{K}$, in order
to work with a fixed value of $\delta$ only. We report in
Fig.~\ref{recombination plot} two plots (for $\delta=0.05$ and
$\delta=0.15$, corresponding to $q=0.9$ and $q=0.7$, respectively)
of the DR plasma rate~\ref{our plasma rate} for $C^{3+}$ (curve 1);
only the 21 resonance levels reported by Mannervik {\it et
al.}~\cite{Man1998} between $0.176\,\mathrm{eV}$ (i.e. $2\cdot
10^{3}\,\mathrm{K}$) and $0.586\,\mathrm{eV}$ (i.e. $7\cdot
10^{3}\,\mathrm{K}$) have been included. Resonance levels above
$0.586\,\mathrm{eV}$ were not taken into account.

As a comparison, the curve 2 represents the experimental fit (Burgess DR rate) of Schippers~{\it et al.}~\cite{Sch2001}, with five fitting levels distributed between $0.169\,\mathrm{eV}$ and $7.969\,\mathrm{eV}$. In Ref.~\cite{Sch2001}, while below $T=10^{4}\,\mathrm{K}$ experimental data are used, above $T=10^{4}\,\mathrm{K}$ the authors use theoretical calculations from the AUTOSTRUCTURE code. The method for obtaining the fit curve 2 is based on converting the laboratory experimental DR rate, as function of the relative energy, into a cross section and convoluting it with an isotropic Maxwellian electron energy distribution.

In curve 3, we report the Maxwellian rate from Eq.~\ref{non-Maxwellian rate}, with $\delta=0$ and $\gamma_{i}\neq 0$; again this Maxwellian curve contains only the 21 levels of Mannervik~{\it et al.}~\cite{Man1998}. This curve is always higher than the curve 2: a possible explanation could be that the strengths of the most important levels, considered in Ref.~\cite{Man1998}, are too large.

The curve 4 is then a new fit that takes into account the deformation of the distribution ($\delta\neq 0$) with $\gamma_{i}=0$ and $\bar{\sigma}_{i}\,E_{\mathrm{res},i}$ proportional to the coefficients $c_{i}$ of Ref.~\cite{Sch2001}. We define it as
\begin{equation}
\alpha_{\mathrm{fit}}^{\mathrm{DR}}(k_{B} T)=\frac{1}{T^{3/2}}\left(1+\frac{15}{4}\delta\right)\sum_{i=1}^{5}c_{i}\exp\left[-\frac{E_{i}}{k_{B} T}-\delta\left(\frac{E_{i}}{k_{B} T}\right)^{2}\right]\, .\label{non-Maxwellian fit}
\end{equation}

The $\alpha_{\mathrm{fit}}^{\mathrm{DR}}(k_{B} T)$ lies inside the 20 per cent of uncertainty of the fit rate in the temperature region considered.

As evident from Fig.~\ref{recombination plot}, the effect of the deformation is to increase the value of the maximum, with a small shift of its position towards higher temperatures, and to lower the rate of curve 1 below the curve 2 for $T<0.8\cdot 10^{3}\,\mathrm{K}$. This is due to the fact that the resonance levels considered are all above this low temperature.

\section{Conclusions}
In conclusion, with the correction introduced due to small deformations of the electron distribution, the calculated corrected DR rate agrees with the experimental fit at low temperature, which is given with an uncertainty of about 20 per cent.

We remark that when using information obtained in laboratory DR experiments on the DR rate to discuss properties of astrophysical plasmas, we must
consider the different state of deformation in the distribution function of the laboratory and the astrophysical plasmas, if both are not in the Maxwellian equilibrium state.


\begin{thebibliography}{99}
\bibitem{Mul1999} A.~M\"{u}ller, Int. Jour. of Mass Spectroscopy~{\bf 192} (1999) 9.\\
\bibitem{Sav2000} N.~R.~Badnell, astro-ph/0604144.\\
\bibitem{Sch2001} S.~Schippers {\it et al.}, Ap.~J.~{\bf 555} (2001) 1027.\\
\bibitem{Sch2006} E.~W.~Schmidt {\it et al.}, Ap.~J.~{\bf 641} (2006) 157, (astro-ph/0603340).
\bibitem{Sho1969} B.~W.~Shore, Ap.~J.~{\bf 158} (1969) 1205.
\bibitem{Col1989} G.~Collins II, The fundamentals of stellar astrophysics, W.~H.~Freeman and Co., New York, (1989) 398.
\bibitem{Guo2004} G.-X.~Chen and A.~K.~Pradhan, astro-ph/0510534.\\
\bibitem{Spi1940} L.~Spitzer Jr., Month. Not. Roy. Astr. Soc.~{\bf 100} (1940) 402.\\
\bibitem{Mae2006} G.~Maero, P.~Quarati and F.~Ferro, Eur.~Phys.~J B {\bf 50} (2006) 23.\\
\bibitem{Fer2005} F.~Ferro and P.~Quarati, Phys. Rev. E {\bf 71}, (2005) 026408.\\
\bibitem{Fer2006} F.~Ferro, F.~Bachari, G.~Maero and P.~Quarati, submitted to PoS, (2006).\\
\bibitem{Boo2005} J.~P.~Boon and C.~Tsallis Eds., Europhys. News~{\bf 36} (2005) 185.\\
\bibitem{Gel2004} M.~Gell-Mann and C.~Tsallis, Nonextensive Entropy-Interdisciplinary Applications, Oxford University Press, New York, 2004.\\
\bibitem{Sav2006} D.~W.~Savin {\it et al.}, Ap.~J.~{\bf 642} (2006) 1275.
\bibitem{Cor1999} M.~Coraddu {\it et al.}, Braz.~J.~Phys.~{\bf 29} (1999) 153, (nucl-th/9811081).
\bibitem{Man1998} S.~Mannervik {\it et al.}, Phys.~Rev.~Lett.~{\bf 81} (1998) 313.
\end{thebibliography}
\end{document}